\documentclass[pra,twocolumn,
superscriptaddress,
longbibliography,
aps
]{revtex4-2}
\usepackage{graphicx}
\usepackage{amsmath}
\usepackage{amsfonts}
\usepackage{amssymb}
\usepackage[colorlinks=true,linkcolor=blue,urlcolor=blue,citecolor=blue]{hyperref}
\usepackage{color}
\usepackage{dsfont}
\usepackage{placeins} 
\usepackage{bm}           

\usepackage{times}

\usepackage{scalerel}
\usepackage{tikz}
\usetikzlibrary{svg.path}
\definecolor{orcidlogocol}{HTML}{A6CE39}
\tikzset{
	orcidlogo/.pic={
		\fill[orcidlogocol] svg{M256,128c0,70.7-57.3,128-128,128C57.3,256,0,198.7,0,128C0,57.3,57.3,0,128,0C198.7,0,256,57.3,256,128z};
		\fill[white] svg{M86.3,186.2H70.9V79.1h15.4v48.4V186.2z}
		svg{M108.9,79.1h41.6c39.6,0,57,28.3,57,53.6c0,27.5-21.5,53.6-56.8,53.6h-41.8V79.1z M124.3,172.4h24.5c34.9,0,42.9-26.5,42.9-39.7c0-21.5-13.7-39.7-43.7-39.7h-23.7V172.4z}
		svg{M88.7,56.8c0,5.5-4.5,10.1-10.1,10.1c-5.6,0-10.1-4.6-10.1-10.1c0-5.6,4.5-10.1,10.1-10.1C84.2,46.7,88.7,51.3,88.7,56.8z};
	}
}
\newcommand\orcid[1]{\href{https://orcid.org/#1}{\mbox{\scalerel*{
				\begin{tikzpicture}[yscale=-1,transform shape]
					\pic{orcidlogo};
				\end{tikzpicture}
			}{R}}}}

\begin{document}
	
	\title{Effective description of cooling and thermal shifts in quantum systems coupled to bosonic modes}
	
	\author{Simon~B.~J\"ager\,\orcid{0000-0002-2585-5246}}
	\email{sjaeger@physik.uni-kl.de}
	\affiliation{Physics Department and Research Center OPTIMAS, Technische Universit\"at Kaiserslautern, D-67663, Kaiserslautern, Germany}
	
	\author{Ralf~Betzholz\,\orcid{0000-0003-2570-7267}}
	\email{ralf\_betzholz@hust.edu.cn}
	\affiliation{School of Physics, International Joint Laboratory on Quantum Sensing and Quantum Metrology, Hubei Key Laboratory of Gravitation and Quantum Physics, Institute for Quantum Science and Engineering, Wuhan National High Magnetic Field Center, Huazhong University of Science and Technology, Wuhan 430074, China}
	
	\begin{abstract}
		Recently, an effective Lindblad master equation for quantum systems whose dynamics are coupled to dissipative bosonic modes was introduced [\href{https://link.aps.org/doi/10.1103/PhysRevLett.129.063601}{Phys. Rev. Lett. \textbf{129} 063601 (2022)}]. In this approach, the bosonic modes are adiabatically eliminated, and one can effectively describe the dynamics of the quantum systems. Here, we demonstrate that this effective master equation can also be used to describe cooling in systems with light-matter interactions. We provide two examples: sideband cooling of an optomechanical oscillator in the unresolved as well as resolved sideband regime and cooling of an interacting quantum system, the transverse-field Ising model. We compare our effective description with a full numerical simulation of the composite formed by the quantum system plus bosonic mode and find excellent agreement. In addition, we present how the effective master equation can be extended to the case of nonvanishing mean thermal occupations of the bosonic mode. We use this approach to calculate modifications of the linewidth and frequency for a two-level system coupled to a dissipative thermal bosonic mode. Here, we highlight that our approach allows for a massive reduction of the underlying Liouville-space dimension.  
	\end{abstract}
	
	\maketitle

	\section{Introduction}
	Engineering the dissipation in a quantum system (QS) is an exciting possibility to cool its degrees of freedom and to stabilize desired correlated quantum states~\cite{Poyatos:1996,Verstraete:2009}. Dissipation can, for example, be engineered by coupling the QS strongly to specific modes of the electromagnetic field that can irreversibly transfer energy from the QS into free space. This strong coupling may be realized by trapping radiation in geometries such as fibers~\cite{Kato:2015}, waveguides~\cite{Gu:2017,Vadiraj:2021,Terradas:2022}, resonators~\cite{Kippenberg:2007,Aspelmeyer:2014}, and cavities~\cite{Walther:2006,Ritsch:2013} that effectively modify the density of states of the electromagnetic field. 
	
	For instance, cavity cooling~\cite{Maunz:2004,Aspelmeyer:2014,Hosseini:2017} is based on this principle, in which the resonance frequency of a cavity is blue detuned from resonances in the QS increasing the probability of the emission of high-energy photons into free space, which leaves the QS at a lower energy due to energy conservation. However, this simplified picture ignores that thermal and quantum back-action of the electromagnetic field can also result in heating of the QS, and therefore, only a thorough analysis of the underlying mechanisms can uncover the entire cooling potential. This is why, in order to faithfully predict this potential, one requires tools which take into account the correct noise terms, dissipation rates, and also the emerging level shifts in the QS. 
	
	Recently, a master equation of the Lindblad form~\cite{Lindblad:1976} was derived which describes the dynamics of the QS coupled to bosonic modes (BM)~\cite{Jaeger:2022}. This master equation describes the effective dissipative dynamics of the QS's density matrix without involving the BM, which allows for an efficient simulation due to a massive reduction of the underlying Liouville-space dimension. This master equation has been proposed as a tool to simulate open many-body QS with engineered interactions and dissipation by the BM. With this, it is a perfect fit to analyze cooling in QS coupled to BM. However, this potential has not been exploited yet and is at the center of this paper. 
	
	More precisely, the purpose of this paper is twofold: (i)~We apply this effective treatment to describe and analyze cooling of the QS. Here, we investigate two specific examples of QSs and, in both cases, compare the effective steady state and dynamics to an exact treatment, be it analytical or numerical. With these two examples, we present evidence that this master equation provides a faithful description of the effective QS dynamics. Furthermore, in each example, we highlight a certain advantage, namely, the validity for a wide range of the QS-BM coupling strength, including strong-coupling effects that are not described by previous treatments, as well as the applicability to interacting many-body QSs. (ii) Under certain conditions, we generalize the effective master equation derived in Ref.~\cite{Jaeger:2022} to the case where the mean thermal occupation of the BM is different from zero, which is of particular interest for scenarios in which the BM frequency lies well below the optical regime. In this case, by means of a paradigm model system, the quantum Rabi model, we show that the effective treatment provides accurate results that agree with exact numerical treatments, in which one requires very high Hilbert-space truncation dimensions, even for moderately low thermal occupation numbers.
	
	This paper is structured as follows. In Sec.~\ref{Sec:me}, we introduce the theoretical model that lies at the basis of our investigation. In Sec.~\ref{Sec:zero_temp}, we present the cooling dynamics described by the effective Lindblad master equation for the zero-temperature case by applying it to two different physical systems in Secs.~\ref{Sec:opto}~and~\ref{Sec:ising}. The finite-temperature case in certain scenarios is shown in Sec.~\ref{Sec:finite_temp}, including an exemplary application in Sec.~\ref{Sec:Rabi}. We conclude our results in Sec.~\ref{Sec:conclusion}. For the sake of completeness, the details of the derivation of the effective master equation are provided in Appendixes~\ref{app:A} and~\ref{app:B}.

	\section{Theoretical model}
	\label{Sec:me}
	In this section, we introduce the class of dissipative setups under investigation and present the effective master equation that we employ to study them. A detailed derivation of such an effective description was shown in Ref.~\cite{Jaeger:2022} for the case of a quasi-zero-temperature environment consisting of multiple BMs. In this paper, however, we will focus on only the single-mode case.
	\begin{figure}[tb]
		\center\includegraphics[width=\linewidth]{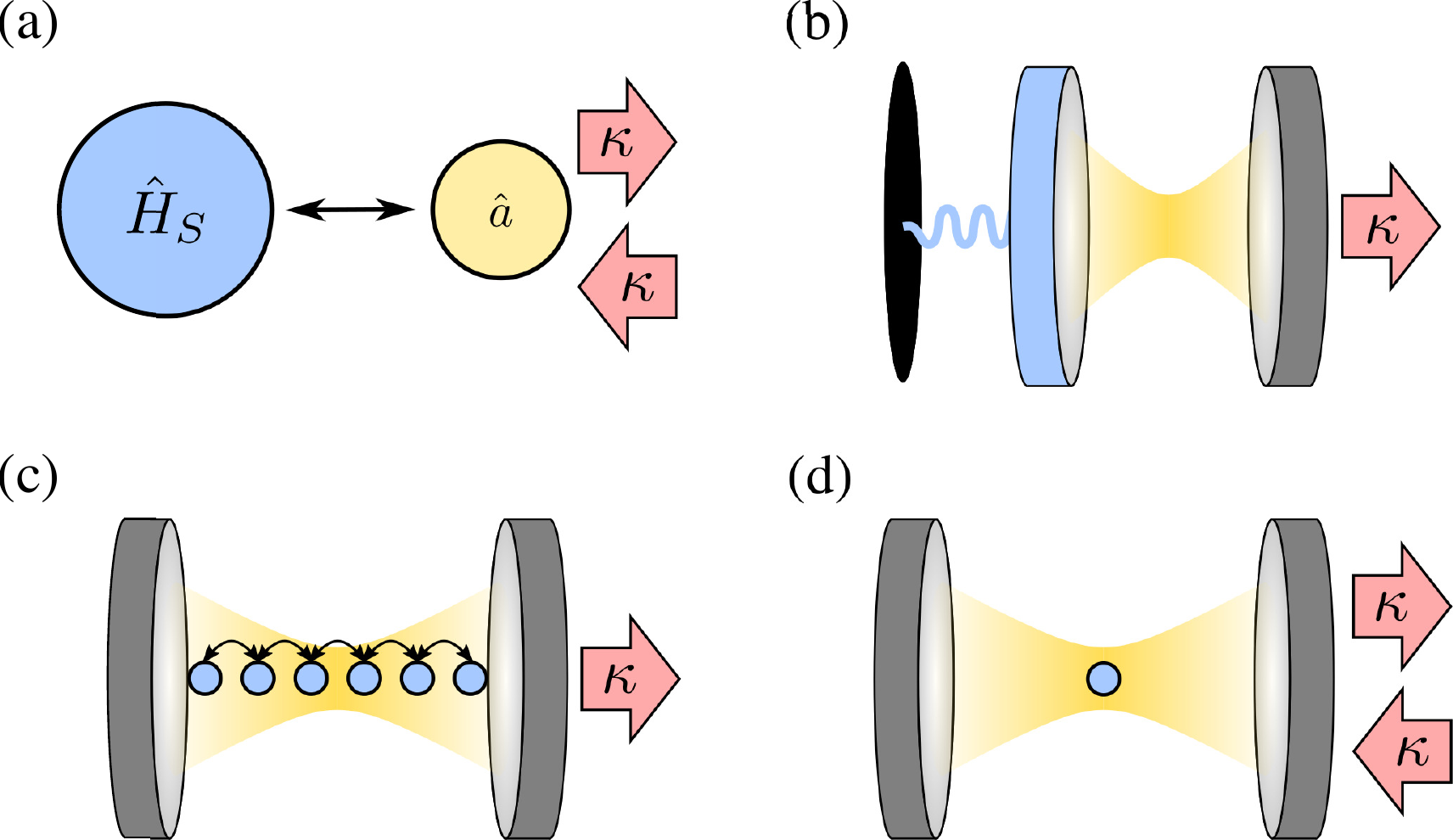}
		\caption{(a) General sketch of the setups that can be treated using the effective master equation~\eqref{eq:Lfull}. (b) A lossy, laser-driven cavity interacting with one of its harmonically moving mirrors via radiation pressure. (c) A transverse-field Ising chain collectively coupling to a cavity mode. (d) A single atom coupled to a bosonic mode at non-zero temperature.
			\label{Fig:1}}
	\end{figure}
	
	We consider the composite of the BM and QS to be described by a Born-Markov master equation 
	\begin{equation}
		\label{eq:Lfull}
		\mathcal{L}\hat{\rho}=-i[\hat{H},\hat{\rho}]+\mathcal{L}_\text{th}\hat{\rho},
	\end{equation}
	where $\hat{\rho}$ is the density matrix of the composite of the BM and QS. Here, the coherent dynamics is governed by the Hamiltonian
	\begin{equation}
		\label{eq:H}
		\hat{H}=\hat{H}_S+\hat{a}^{\dag}\hat{\Omega}_S\hat{a}+\hat{a}^{\dag}\hat{S}+\hat{S}^{\dag}\hat{a}, 
	\end{equation}
	where we have introduced the BM annihilation operator $\hat{a}$. 
	Furthermore, $\hat{H}_S$ denotes the QS Hamiltonian in the absence of the BM. The remaining terms describe the coupling between the QS and the BM. In detail, the term including $\hat{\Omega}_S$ represents the BM energy. Due to the presence of the QS, it can possibly include operators of the QS degrees of freedom. The last part, $\hat{a}^{\dag}\hat{S}+\hat{S}^{\dag}\hat{a}$, is the BM driving term, which may likewise contain QS operators in $\hat{S}$.
	
	In addition to the coherent part, we assume the BM is in contact with a thermal environment at inverse temperature $\beta$. In a Born-Markov approximation, its thermalization with the environment is described by the superoperator
	\begin{equation}
		\label{eq:Ld}
		\mathcal{L}_\text{th}=\kappa \{(\bar{n}+1)\mathcal{D}[\hat{a}]+\bar{n}\mathcal{D}[\hat{a}^\dagger]\}
	\end{equation}
	acting on the density operator describing the BM degrees of freedom. The action of the dissipators $\mathcal{D}$ is given by $\mathcal{D}[\hat{O}]\hat{\rho}=2\hat{O}\hat{\rho}\hat{O}^{\dag}-\hat{O}^{\dag}\hat{O}\hat{\rho}-\hat{\rho}\hat{O}^{\dag}\hat{O}$. Furthermore, the thermalization rate is denoted by $\kappa$ and the mean thermal occupation number of the BM at the environment temperature is $\bar{n}=[\exp(\beta\omega_c)-1]^{-1}$, with $\omega_c$ being the BM resonance frequency.
	
	The Supplemental Material of Ref.~\cite{Jaeger:2022} shows that for a quasi-zero-temperature environment ($\bar{n}\approx 0$), the time evolution of the reduced density matrix of the QS, $\hat{\rho}_{\mathrm{sys}}=\mathrm{Tr}_{\mathrm{BM}}(\hat{D}^{\dag}\hat{\rho}\hat{D})$, can be decoupled up to second order in $||\hat{\alpha}||$ and $\hat{S}$ under the transformation 
	\begin{equation}\label{eq:D}
		\hat{D}=\exp(\hat{\alpha}^\dagger \hat{a}-\hat{a}^\dagger \hat{\alpha}).
	\end{equation}
	Physically, this decoupling is based on a weak-coupling approximation between the BM and the QS.
		Above, $\mathrm{Tr}_{\mathrm{BM}}(\,\cdot\,)$ denotes the trace over the BM degrees of freedom and the operator $\hat{\alpha}$ has to be determined for every specific setup at hand. In fact, in the original picture, before the displacement operation is performed, $\hat{\alpha}$ can be seen as the partial trace over the BM operator $\hat{a}$, i.e., $\hat{\alpha}\hat{\rho}_{\mathrm{sys}}\approx\mathrm{Tr}_{\mathrm{BM}}(\hat{a}\hat{\rho})$.
	
	In the next two sections, we will separately analyze the case of a zero bosonic occupation and a finite bosonic occupation and, in both cases, show how the decoupling can be achieved by a respectively appropriate choice of the effective-field operator $\hat{\alpha}$ that solves the corresponding elimination condition. We will also give some examples to demonstrate the applicability of this method in both cases. In order to do so, we will compare the simulation of the effective master equations to numerical treatments of the full master equations without the elimination of the BM.

	\section{Effective Lindblad master equation---zero bosonic occupation}
	\label{Sec:zero_temp}
	Let us first focus on the zero bosonic occupation, $\bar{n}=0$, that was introduced in Ref.~\cite{Jaeger:2022}. In order to effectively decouple the QS from the BM in this situation, the condition that the effective-field operator $\hat\alpha$ has to fulfill reads
	\begin{equation}
		\frac{\partial\hat{\alpha}}{\partial t}=-i[\hat{H}_S,\hat{\alpha}]-i\hat{\Omega}_S\hat{\alpha}-i\hat{S}-\kappa\hat{\alpha}.\label{eq:alpha}
	\end{equation}
	Once the solution $\hat\alpha$ that solves this condition is found, the Liouvillian governing the effective master equation $\partial\hat{\rho}_\text{sys}/\partial t=\mathcal{L}_\text{eff}\hat\rho_{\text{sys}}$ of the QS is given by
	\begin{equation}
		\label{eq:Leff_zero}
		\mathcal{L}_{\text{eff}}\hat{\rho}_{\text{sys}}=-i[\hat{H}_{\text{eff}},\hat{\rho}_{\text{sys}}]+\kappa\mathcal{D}[\hat{\alpha}]\hat{\rho}_{\text{sys}},
	\end{equation}
	where the coherent dynamics is described by the effective Hamiltonian
	\begin{equation}
		\hat{H}_{\mathrm{eff}}=\hat{H}_S+\frac{1}{2}(\hat{\alpha}^{\dag}\hat{S}+\hat{S}^{\dag}\hat{\alpha}).\label{eq:Heff_zero}
	\end{equation}
	
	In summary, solving the elimination condition~\eqref{eq:alpha} yields the operator $\hat{\alpha}$ for a given set of QS operators $\hat{H}_S$, $\hat{\Omega}_S$, and $\hat{S}$. In the effective dynamics of the QS degrees of freedom, this $\hat{\alpha}$ then appears both in an effective QS driving term and in the dissipator of the originally non-dissipating QS, which is induced by the coupling to the BM. 
	
	\subsection{Cavity cooling of a mechanical oscillator}
	\label{Sec:opto}
	As a first example, we will now use this approach to describe cooling in an optomechanical setup [see Fig.~\ref{Fig:1}(b)]. Here, the BM is a single mode of a laser-driven, lossy optical cavity with linewidth $\kappa$ that is interacting with one of its harmonically suspended mirrors via the radiation-pressure force~\cite{Aspelmeyer:2014,Wilson-Rae:2007,Marquardt:2007}. Specifically, we consider the QS Hamiltonian to be
	\begin{equation}
		\hat{H}_S= \omega_0 \hat{b}^{\dag}\hat{b},\label{HSoptomech}
	\end{equation}
	where $\omega_0$ and $\hat b$ respectively denote the bare frequency and the annihilation operator of the mirror motion. The coupling to the driven cavity, on the other hand, is described by
	\begin{gather}
		\hat{\Omega}_S=-\Delta+g(\hat{b}^{\dag}+\hat{b}),\label{OmegaSoptomech}\\
		\hat{S}=\eta,\label{Soptomech}
	\end{gather}
	with the detuning $\Delta=\omega_\text{L}-\omega_c$ between the driving laser of frequency $\omega_\mathrm{L}$ and the cavity, $\eta$ being the driving strength, and $g$ being the optomechanical coupling strength.
	
	Since we are considering optical cavities, the restriction to the case of $\bar{n}=0$ is a reasonable assumption. The elimination condition~\eqref{eq:alpha} for $\hat\alpha$ at steady state, which we need to solve to obtain our effective description, then reads
	\begin{equation}
		\omega_0[\hat b^\dagger \hat b,\hat\alpha]+g(\hat b+\hat b^\dagger)\hat\alpha-\left(\Delta+i\kappa\right)\hat\alpha+\eta=0.
	\end{equation}
	To solve this equation for $\hat{\alpha}$, it is helpful to introduce the operator $\tilde\alpha=\hat{T}\hat\alpha$~\cite{Torres:2019,Betzholz:2020}, with the translation operator ${\hat{T}=\exp[g(\hat{b}^\dagger-\hat{b})/\omega_0]}$. This transforms the above equation into 
	\begin{equation} 
		\omega_0 [\tilde\alpha,\hat b^\dagger \hat b]+\left(\Delta+i\kappa+\frac{g^2}{\omega_0}\right)\tilde\alpha=\eta \hat{T}.
	\end{equation}
	Expanding this in the mechanical-oscillator Fock states $|m\rangle$, with $\hat b|m\rangle=\sqrt{m}|m\rangle$, and applying $\hat T^\dagger$ from the left then lets us arrive at
	\begin{equation}
		\hat\alpha=\eta \hat{T}^\dagger\sum_{m=0}^\infty\left[(m-\hat b^\dagger \hat b)\omega_0+\Delta+i\kappa+\frac{g^2}{\omega_0}\right]^{-1}\hat{T} \vert m\rangle\langle m\vert
	\end{equation}
	for the original operator $\hat{\alpha}$. Executing the remaining translation transform on the right-hand side finally yields 
	\begin{equation}
		\label{eq:alphaoptomech}
		\hat{\alpha}=\sum_{m=0}^\infty\frac{\eta}{\Delta+m\omega_0+i\kappa-\omega_0 \hat{b}^\dagger\hat{b}-g(\hat{b}+\hat{b}^{\dag})}  \vert m\rangle\langle m\vert
	\end{equation}
	which is then used to establish the effective Hamiltonian~\eqref{eq:Heff} and Liouvillian~\eqref{eq:Leff_zero}.
	
	Such optomechanical setups show cooling in the regime $\Delta<0$~\cite{Marquardt:2007,Aspelmeyer:2014}. In order to show that our effective model also accurately reproduces this cooling in the weak-coupling regime $g\ll|\Delta+i\kappa|$ we can expand $\hat{\alpha}$, 
	according to~\footnote{We use the series expansion $(q-c x)^{-1}=q^{-1}\sum_{k=0}^\infty(cx/q)^k$}, leading to
	\begin{equation}
		\hat{\alpha}=\sum_{m,k=0}^\infty \frac{\eta }{\Delta+m\omega_0+i\kappa}\left[\frac{\omega_0 \hat{b}^\dagger \hat{b}+g(\hat{b}+\hat{b}^\dagger)}{\Delta+m\omega_0+i\kappa}\right]^k
		\vert m\rangle\langle m\vert. \label{eq:alphaold}
	\end{equation}
	In the weak-coupling regime, $g\ll\kappa$, we expand $\hat{\alpha}$ up to first order in $g/\kappa$ such that
		\begin{align}
			&[\omega_0 \hat{b}^\dagger \hat{b}+g(\hat{b}+\hat{b}^\dagger)]^k\vert m\rangle \nonumber\\
			&\approx\omega_0m^k|m\rangle+c_+(m)|m+1\rangle+c_{-}(m)|m-1\rangle\label{eq:hilf}
		\end{align}
		with
		\begin{align}
			c_{-}(m)=&\omega_0^{k-1}g\sqrt{m}\sum_{l=0}^{k-1}(m-1)^lm^{k-1-l}\nonumber\\
			=&\omega_0^{k-1}g\sqrt{m}[m^k-(m-1^k)],\\
			c_{+}(m)=&\omega_0^{k-1}g\sqrt{m+1}\sum_{l=0}^{k-1}(m+1)^lm^{k-1-l}\nonumber\\
			=&\omega_0^{k-1}g\sqrt{m+1}[(m+1)^k-m^k].
		\end{align}
		Using these results and inserting them in Eq.~\eqref{eq:alphaold} we can use the geometric series $\sum_{k=0}^\infty x^k=1/(1-x)$, with $x=m\omega_0/(\Delta+m\omega_0+i\kappa)$ or $x=(m\pm1)\omega_0/(\Delta+m\omega_0+i\kappa)$, to arrive at the closed expression
	\begin{equation}
		\label{eq:alphaoptomechapp}
		\hat{\alpha}=\frac{\eta}{\Delta+i\kappa}\left[1+\left(\frac{g\hat{b}}{\Delta+\omega_0+i\kappa}+\frac{g\hat{b}^{\dag}}{\Delta-\omega_0+i\kappa}\right)\right].
	\end{equation}
	If we further assume that the frequency $\omega_0$ is sufficiently large, such that we can neglect all counter-rotating terms in the effective master equation~\eqref{eq:Leff_zero}, we may approximate it as
	\begin{equation}
		\label{eq:simplyeffoptoMaster}
		\mathcal{L}_\text{eff}\hat{\rho}_\text{sys}\approx-i\omega_0[\hat{b}^\dagger\hat{b},\hat{\rho}_\text{sys}]+A_-\mathcal{D}[\hat{b}]\hat{\rho}_{\text{sys}}+A_+\mathcal{D}[\hat{b}^{\dag}]\hat{\rho}_\text{sys},
	\end{equation}
	where we have introduced the cooling rate $A_-$ and the heating rate $A_+$, which have the form
	\begin{equation}
		A_{\mp}=\frac{\kappa g^2\eta^2}{(\Delta^2+\kappa^2)[(\Delta\pm\omega_0)^2+\kappa^2]}.
	\end{equation}
	From these rates it is straightforward to calculate the mean number of motional excitations $\bar{m}=\langle\hat{b}^{\dag}\hat{b}\rangle$ at steady state, which reads
	\begin{equation}
		\label{eq:napprox}
		\bar{m}=\frac{A_{+}}{A_{-}-A_{+}}=\frac{(\Delta+\omega_0)^2+\kappa^2}{-4\Delta\omega_0}
	\end{equation}
	and coincides with the result reported in Ref.~\cite{Wilson-Rae:2007}. This mean excitation number exhibits a minimum at $\Delta=-\sqrt{\kappa^2+\omega_0^2}$. In the non-resolved sideband regime, $\kappa\gg\omega_0$, the minimum is achieved at $\Delta\approx-\kappa$, whereas in the resolved sideband regime,  $\kappa\ll\omega_0$, it is achieved at $\Delta\approx-\omega_0$.
	
\begin{figure}[tb]
		\flushleft\normalsize{(a)}\hspace{-2.4ex}\includegraphics[width=1\linewidth]{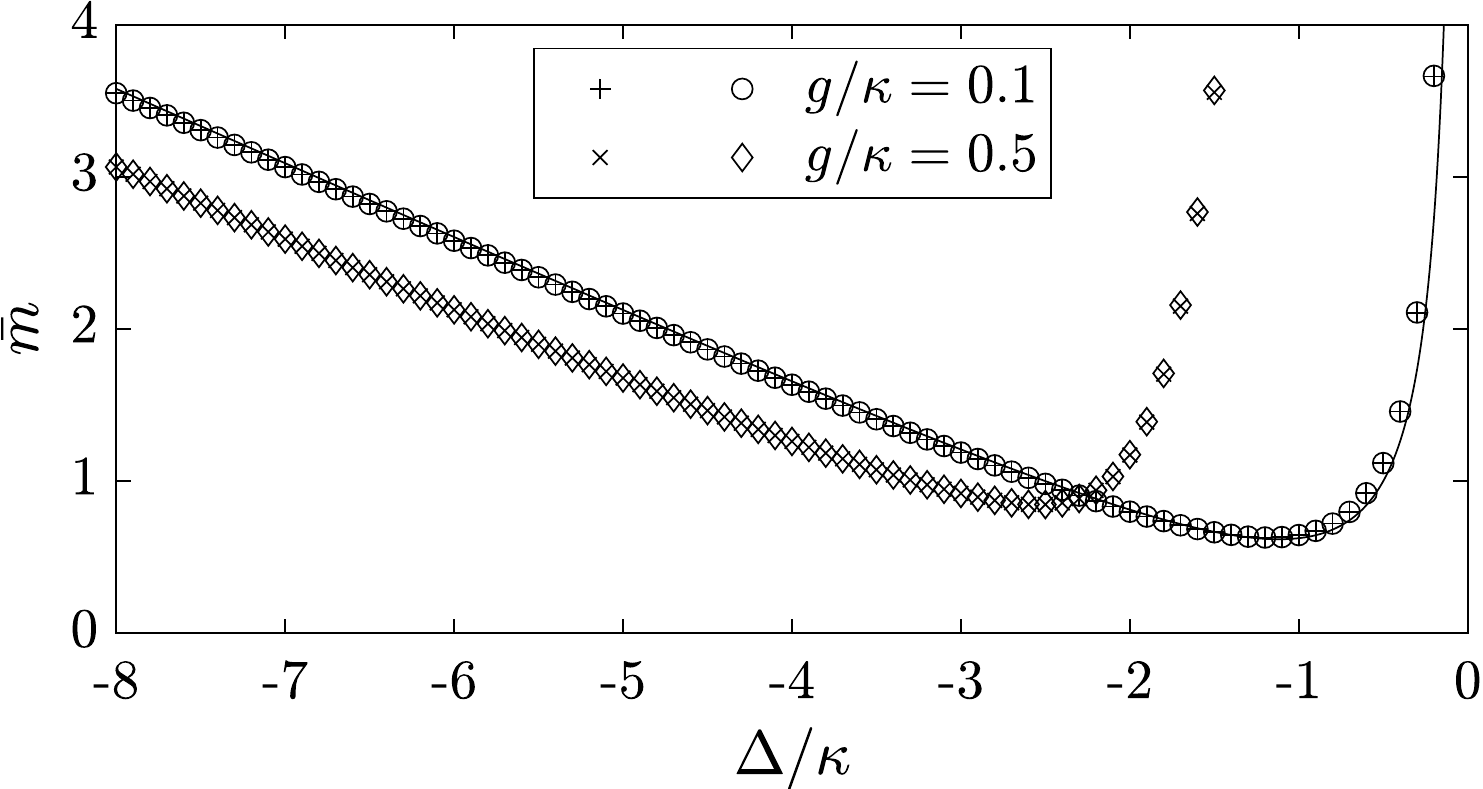}\vspace{-3ex}
		\flushleft\normalsize{(b)}\hspace{-2.4ex}\includegraphics[width=\linewidth]{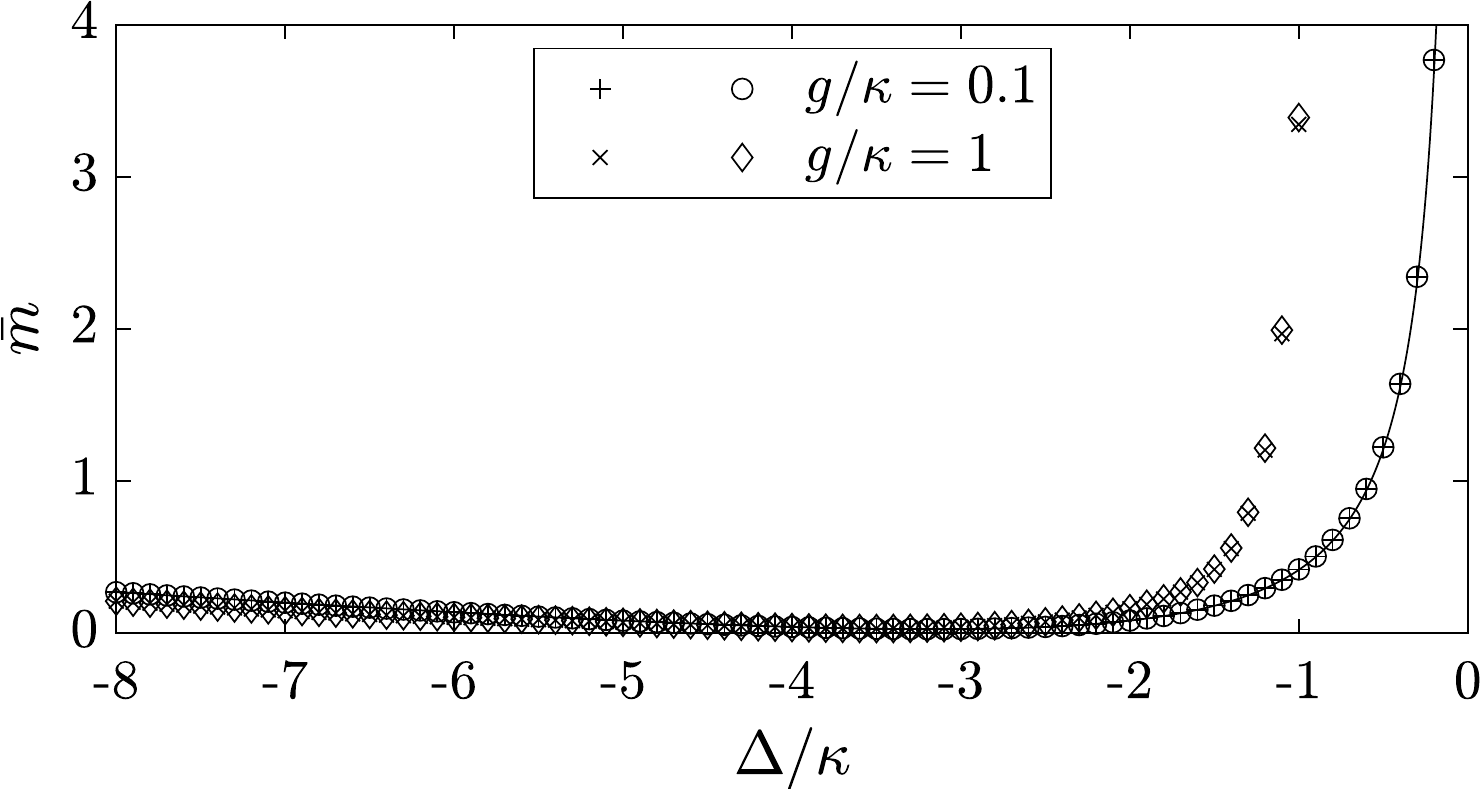}
		\caption{Mean excitation number $\bar{m}=\langle\hat{b}^{\dag}\hat{b}\rangle$ at steady state as a function of $\Delta$ in units of $\kappa$ for (a) $\omega_0/\kappa=0.5$ and (b) $\omega_0/\kappa=3$. Pluses and crosses correspond to the steady state of the effective master equation~\eqref{eq:Leff_zero}  with Eq.~\eqref{eq:alphaoptomech} for different values of $g/\kappa$ (see insets). Circles and diamonds show the steady state of the full master equation~\eqref{eq:Lfull}. The dashed lines represent the approximate result~\eqref{eq:napprox}. In both cases, the driving strength is $\eta/\kappa=0.1$. \label{Fig:2}}
	\end{figure}

	We compare these known approximate results with the ones obtained by calculating the steady state of the full master equation~\eqref{eq:Lfull} and the effective master equation~\eqref{eq:Leff_zero}, in which we use Eq.~\eqref{eq:alphaoptomech}. In Fig.~\ref{Fig:2}, we show the mean number of excitations $\bar{m}$ [Eq.~\eqref{eq:napprox}] at steady state as a function of the detuning $\Delta$ in units of $\kappa$ as solid lines. The pluses and crosses correspond to the steady state of the effective master equation obtained for the two values of $g/\kappa$ that are visible in the legends of Fig.~\ref{Fig:2}(a) and~\ref{Fig:2}(b). For comparison, the results of the full master equation are shown as circles and diamonds for the same values of $g/\kappa$. In principle, when calculating expectation values of system oberservables, such as $\hat{b}^\dagger\hat{b}$, one has to take into account corrections originating from the displacement $\hat{D}$ when going to the transformed picture. The lowest-order correction is of second order in $\hat{\alpha}$, and we found it to be negligible in all presented results.

	For all parameters, we find very good agreement of the results obtained from the full master equation and the results calculated from the effective master equation. The approximate result~\eqref{eq:napprox} for $\bar{m}$, which was derived in the weak coupling regime, is shown by solid lines and is, again, in good agreement for $g/\kappa=0.1$. However, for larger ratios $g/\kappa$, for both $\omega_0/\kappa=0.5$ [Fig.~\ref{Fig:2}(a)] and $\omega_0/\kappa=3$ [Fig.~\ref{Fig:2}(b)], we find significant discrepancies between the results of Eq.~\eqref{eq:napprox} and the steady state result of the effective and full master equation. This highlights that the approximate result is valid in only the weak-coupling regime, while the effective description is still accurate beyond the often assumed weak-coupling approximation.

	\subsection{Cavity cooling of an Ising chain}
	\label{Sec:ising}
	As a second case study, we will show that the effective master equation~\eqref{eq:Leff_zero} can also be used to describe cooling of an interacting many-body QS. In order to do so, we study the dynamics of a transverse-field Ising model~\cite{Elliot:1970}, with open boundary conditions, that is coupled to a single-mode cavity via a Jaynes-Cummings interaction [see Fig.~\ref{Fig:1}(c)]. The QS Hamiltonian in this case is given by
	\begin{equation}
		\hat{H}_S=h\sum_{n=1}^N\hat{\sigma}_n^z-J\sum_{n=1}^{N-1}\hat{\sigma}_n^x\hat{\sigma}_{n+1}^x,\label{HSIsing}
	\end{equation}
	where $h$ denotes the transverse field, $J$ is the nearest-neighbor interaction, and $\hat{\sigma}^q_n$, for $q\in \{x,y,z\}$ and $n=1,\dots,N$, represent the Pauli operators of the $n$th atom. The coupling of the spins to the cavity field, on the other hand, is described by
	\begin{gather}
		\hat{\Omega}_S=\omega_c,\label{OmegaSIsing}\\
		\hat{S}=g\sum_{n=1}^N\hat{\sigma}_n^{-},\label{SIsing}
	\end{gather}
	with $\omega_c$ being the cavity frequency, $g$ being the Jaynes-Cummings interaction strength, and $\hat{\sigma}_n^\pm=(\hat{\sigma}_n^x\pm i\hat{\sigma}_n^y)/2$. Similar models have been used to study the interplay between matter-matter and light-matter interactions~\cite{Mazza:2019, Rohn:2020}. Here, however, we are interested in the dissipative, or more precisely, the cooling dynamics of the spins.
	
	Following the idea presented in Ref.~\cite{Raghunandan:2020}, where a dissipative ancillary spin at the end of the chain is leveraged to sympathetically cool the Ising chain into the ground-state manifold, in our case, we can employ the dissipation of the cavity to cool the many-body QS in the same spirit. 
	
	To achieve efficient cooling, we choose the cavity frequency $\omega_c$ to be equal to the splitting between the ground-state energy and the first-excited-state energy of the Ising chain. Furthermore, we impose that the cavity linewidth must be able to resolve this gap, i.e., $\kappa<\omega_c$. This energy-matching condition then allows to efficiently transfer energy from the spins into the cavity, from which it can be dissipated.
	
\begin{figure}[tb]
		\flushleft\normalsize{(a)}\hspace{-2.4ex}\includegraphics[width=\linewidth]{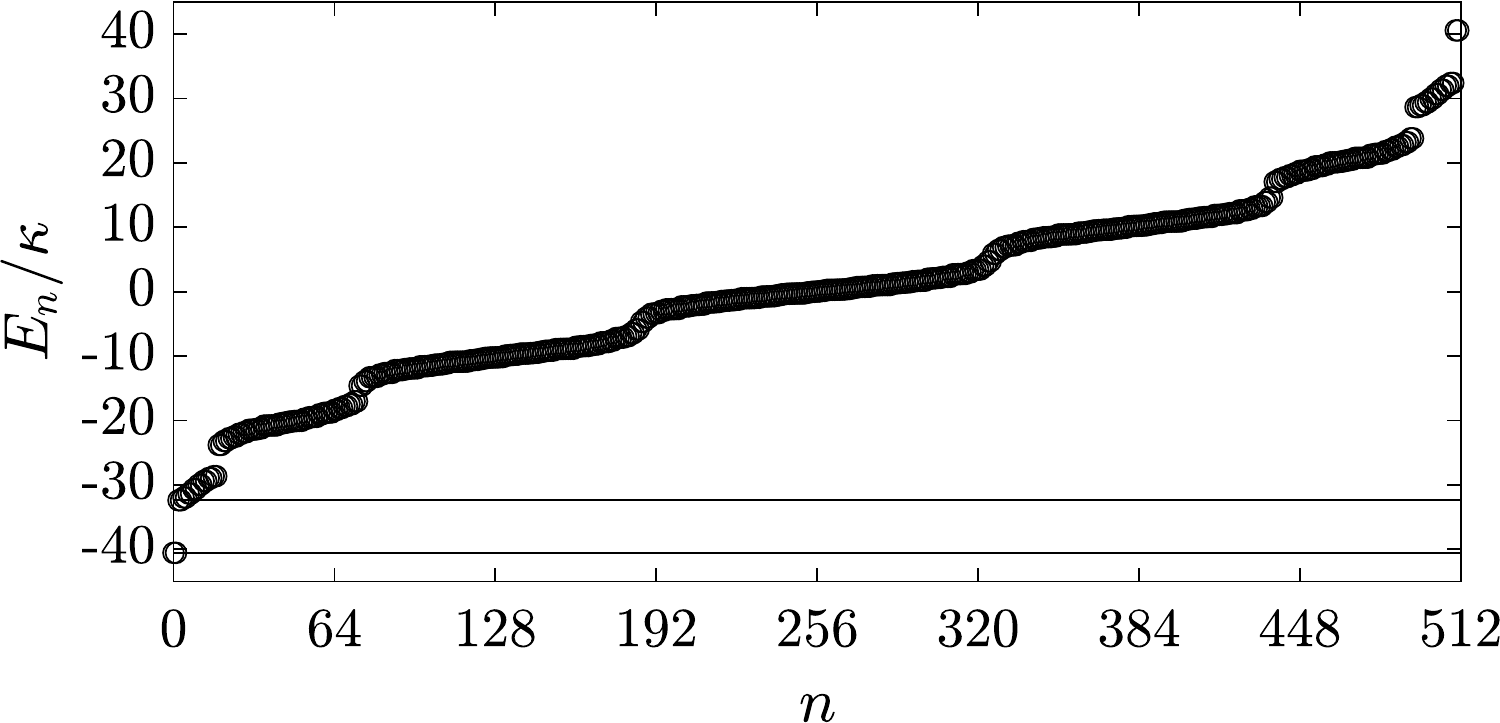}\vspace{-3ex}
		\flushleft\normalsize{(b)}\hspace{-2.4ex}\includegraphics[width=\linewidth]{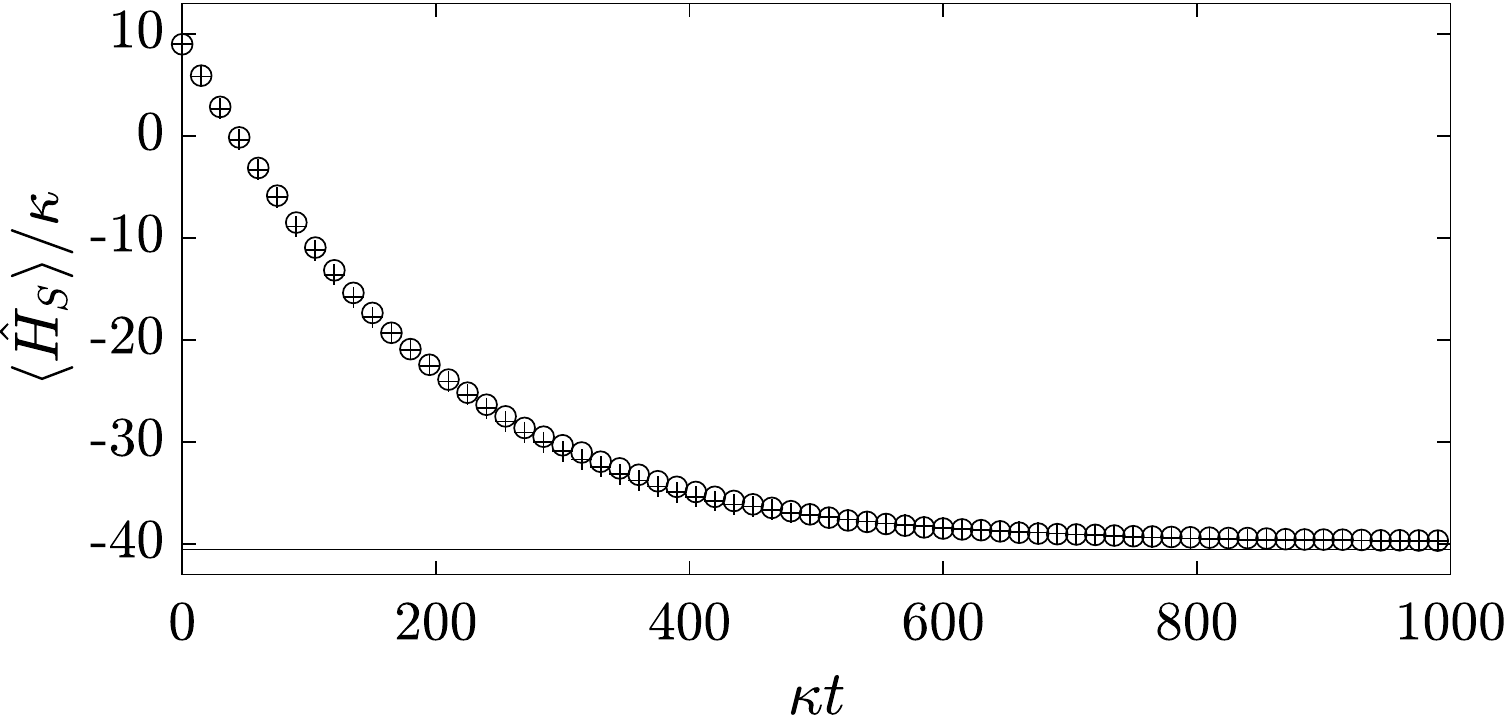}
		\caption{(a) Eigenenergies $E_n$ in units of $\kappa$ of a transverse-field Ising chain with $N=9$, $h/\kappa=1$, and $J/\kappa=5$. The ground-state and first-excited-state energies are shown by the lower and upper horizontal line, respectively. (b) Mean energy $\langle \hat H_S \rangle$ in units of $\kappa$ as a function of time in units of $1/\kappa$ for the same parameters as in (a) and a Jaynes-Cummings interaction strength $g/\kappa=0.3$. The initial state is the fully polarized state with $\langle\hat{\sigma}^z_j\rangle=1$ for all $j$. The horizontal line represents the ground state energy. Pluses are calculated with the effective master equation, and circles are calculated with the full master equation including the cavity field. 
			\label{Fig:3}}
	\end{figure}

	Figure~\ref{Fig:3}(a) depicts the eigenenergies $E_n$ of $\hat H_S$ in units of $\kappa$ for $N=9$, $h/\kappa=1$, and $J/\kappa=5$. The horizontal lines represent the energy splitting between the doubly-degenerate ground state and the first excited state. As mentioned above, for ground-state cooling we set the cavity frequency $\omega$ to match this splitting and calculate the time evolution of the mean energy $\langle \hat H_S \rangle$ of the Ising chain, where we choose the state in which all spins of the chain are in their up state as the initial state. The cooling is shown in Fig.~\ref{Fig:3}(b) for a Jaynes-Cummings interaction strength $g/\kappa=0.3$. Here, circles represent the result obtained using the full master equation~\eqref{eq:Lfull}, and pluses show the one obtained using the effective master equation~\eqref{eq:Leff_zero}. For the calculation of the effective master equation we numerically solve Eq.~\eqref{eq:alpha} for the steady state of $\hat\alpha$ using Eqs.~\eqref{HSIsing}--\eqref{SIsing}. The numerically found operator $\hat{\alpha}$ is then used to calculate the effective Hamiltonian in Eq.~\eqref{eq:Heff_zero} and the effective master equation~\eqref{eq:Leff_zero}. The latter is then used to time evolve the reduced density matrix. The corresponding time evolution is in good agreement with the one obtained from the full master equation~\eqref{eq:Lfull} and shows an efficient cooling of the Ising chain by the dissipative cavity into its ground-state manifold.

	\section{Effective Lindblad master equation---finite temperature}
	\label{Sec:finite_temp}
	In this section, we now turn toward the situation where the BM has a non-vanishing thermal occupation ($\bar{n}\neq0$) at steady-state. While the contribution of a thermal steady-state occupation is negligible for optical frequencies, it can play an important role in the microwave regime if one considers temperatures $T\sim 1$~K or higher. In this regime, it is important to include the effects of the thermal occupation in the QS which we have neglected in the previous section.  The general idea to derive a master equation in the thermal regime is basically the same: we apply the displacement $\hat{D}$ [Eq.~\eqref{eq:D}] to the full master equation [Eq.~\eqref{eq:Lfull}] such that the BMs are, to good approximation, in a thermal state,
	\begin{equation}\label{eq:thermal}
		\hat{\rho}_\text{th}=\frac{1}{\bar{n}+1}\left(\frac{\bar{n}}{\bar{n}+1}\right)^{\hat{a}^\dagger\hat{a}}
	\end{equation}
	after the transformation. Subsequently, we calculate the best choice of $\hat{\alpha}$ that decouples the thermal BM from the QS. In Appendix~\ref{app:A}, we demonstrate this calculation and find that the same calculation is possible if one can treat the commutator $[\hat{\Omega}_S,\hat{\alpha}]$ perturbatively. The reason for this finding is that $\hat{\rho}_\mathrm{th}$ can be a mixture of many Fock states of the BM which can exhibit different frequencies due to $\hat{\Omega}_S$ depending on the state of the QS. Assuming $[\hat{\Omega}_S,\hat{\alpha}]\approx 0$ means that this difference in frequencies is negligible. In this case we find the same expression, Eq.~\eqref{eq:alpha}, that decouples the BM from the QS. It is shown in Appendix~\ref{app:B}, that the resulting Liouvillian that governs the QS's effective master equation $\partial\hat{\rho}_\text{sys}/\partial t=\mathcal{L}_\text{eff}\hat\rho_{\text{sys}}$ is, however, modified and takes the form
	\begin{equation}
		\label{eq:Leff}
		\mathcal{L}_{\text{eff}}\hat{\rho}_{\text{sys}}=-i[\hat{H}_{\text{eff}},\hat{\rho}_{\text{sys}}]+\kappa\{(\bar{n}+1)\mathcal{D}[\hat{\alpha}]+\bar{n}\mathcal{D}[\hat{\alpha}^\dagger]\}\hat{\rho}_{\text{sys}},
	\end{equation}
	where the coherent dynamics is now described by the effective Hamiltonian
	\begin{equation}
		\hat{H}_{\mathrm{eff}}=\hat{H}_S+\frac{1}{2}(\hat{\alpha}^{\dag}\hat{S}+\hat{S}^{\dag}\hat{\alpha})+\frac{\bar{n}}{2}([\hat{\alpha}^{\dag},\hat{S}]+[\hat{S}^{\dag},\hat{\alpha}])+\bar{n}\hat{\Omega}_S.\label{eq:Heff}
	\end{equation}
	
	We will now discuss the additional terms that appear in Eq.~\eqref{eq:Leff} compared to Eq.~\eqref{eq:Leff_zero}, which are identical in the case with $\bar{n}=0$.
	The dissipators in Eq.~\eqref{eq:Leff} are modified because $\bar{n}\neq0$ (in the form of an enhancement of the dissipation rate) and also because of the appearance of the conjugated field operator $\hat{\alpha}^\dag$ as a jump operator. In the effective Hamiltonian~\eqref{eq:Heff}, we find two additional terms that are present only for  $\bar{n}\neq0$. The term proportional to $\hat{\Omega}_S$ gives rise to an additional potential term for the QS originating from the thermal occupation. The terms proportional to $\hat{S}$ are modified for $\bar{n}\neq0$ and describe the influence of thermal fluctuations in emission-absorption and absorption-emission processes of quanta in the BM. In order to bring out the effects of finite $\bar{n}$ clearly, we will analyze a minimal model which consists of a single two-level system which couples to a thermal BM.
	
	\subsection{Dissipative quantum Rabi model at finite temperature}
	\label{Sec:Rabi}
	We now apply the method presented above to a paradigm model of light-matter interaction in the dipole approximation, namely, the quantum Rabi model~\cite{Rabi:1937, Xie:2017}. This model describes a single atomic dipole interacting with the electric field of a single cavity mode [see Fig.~\ref{Fig:1}(b)]. In this case, we identify the QS Hamiltonian
	\begin{equation}
		\hat{H}_S=\frac{\omega_0}{2}\hat{\sigma}^z,
	\end{equation}
	with the atomic transition frequency $\omega_0$, and 
	\begin{gather}
		\hat{\Omega}_S=\omega_c,\\
		\hat{S}=g\hat{\sigma}^x
	\end{gather}
	with $\omega_c$ being the cavity frequency and $g$ being the coupling strength. Here, $\hat{\sigma}^q$, for $q\in \{x,y,z\}$, denote the Pauli operators. For this model, neglecting the commutator $[\hat{\Omega}_S,\hat{\alpha}]$ is exact since $\hat{\Omega}_S=\omega_c$ is a scalar. The steady-state elimination condition~\eqref{eq:alpha} then reads
	\begin{equation}
		\frac{\omega_0}{2}[\hat{\sigma}^z,\hat{\alpha}]+(\omega_c-i\kappa)\hat{\alpha}+g\hat{\sigma}^x=0,
	\end{equation}
	and is readily solved by the effective-field operator
	\begin{equation}
		\hat{\alpha}=\alpha_+\hat{\sigma}^++\alpha_-\hat{\sigma}^-,
	\end{equation}
	where $\alpha_\pm=-g/(\omega_c\pm\omega_0-i\kappa)$ and $\hat{\sigma}^\pm=(\hat{\sigma}^x\pm i\hat{\sigma}^y)/2$. With this, we can derive the effective Hamiltonian in Eq.~\eqref{eq:Heff}, which takes the form
	\begin{equation}
		\hat{H}_{\mathrm{eff}}=\frac{\omega_0+\Delta\omega_0}{2}\hat{\sigma}^z+\frac{\Sigma\omega_0}{2},
  	\end{equation}
	with the frequency shift
	\begin{align}
		\Delta\omega_0=g(2\bar{n}+1)\left[\mathrm{Re}(\alpha_-)-\mathrm{Re}(\alpha_+)\right]
	\end{align}
	and the energy offset
	\begin{align}
		\Sigma\omega_0=g\left[\mathrm{Re}(\alpha_-)+\mathrm{Re}(\alpha_+)\right].
	\end{align}
	The above frequency shift is split into two components, which correspond to the co-rotating terms $\propto\mathrm{Re}(\alpha_-)$ and the counter-rotating terms $\propto\mathrm{Re}(\alpha_+)$. Explicitly, it is given by
	\begin{equation}
		\Delta\omega_0=-(2\bar{n}+1)\left[\frac{g^2(\omega_c-\omega_0)}{(\omega_c-\omega_0)^2+\kappa^2}-\frac{g^2(\omega_c+\omega_0)}{(\omega_c+\omega_0)^2+\kappa^2}\right]
	\end{equation}
	and is proportional to $2\bar{n}+1$. This includes contributions from both thermal excitations and vacuum fluctuations. The co- and counter-rotating terms correspond to the Lamb shift and the Bloch-Siegert shift, respectively. In addition to the effective Hamiltonian, we also calculate using Eq.~\eqref{eq:alpha} the dissipator in Eq.~\eqref{eq:Leff}, thereby forming the effective Liouvillian $\mathcal{L}_\text{eff}$.

	\begin{figure}[tb]
		\flushleft\normalsize{(a)}\hspace{-2.4ex}\includegraphics[width=1\linewidth]{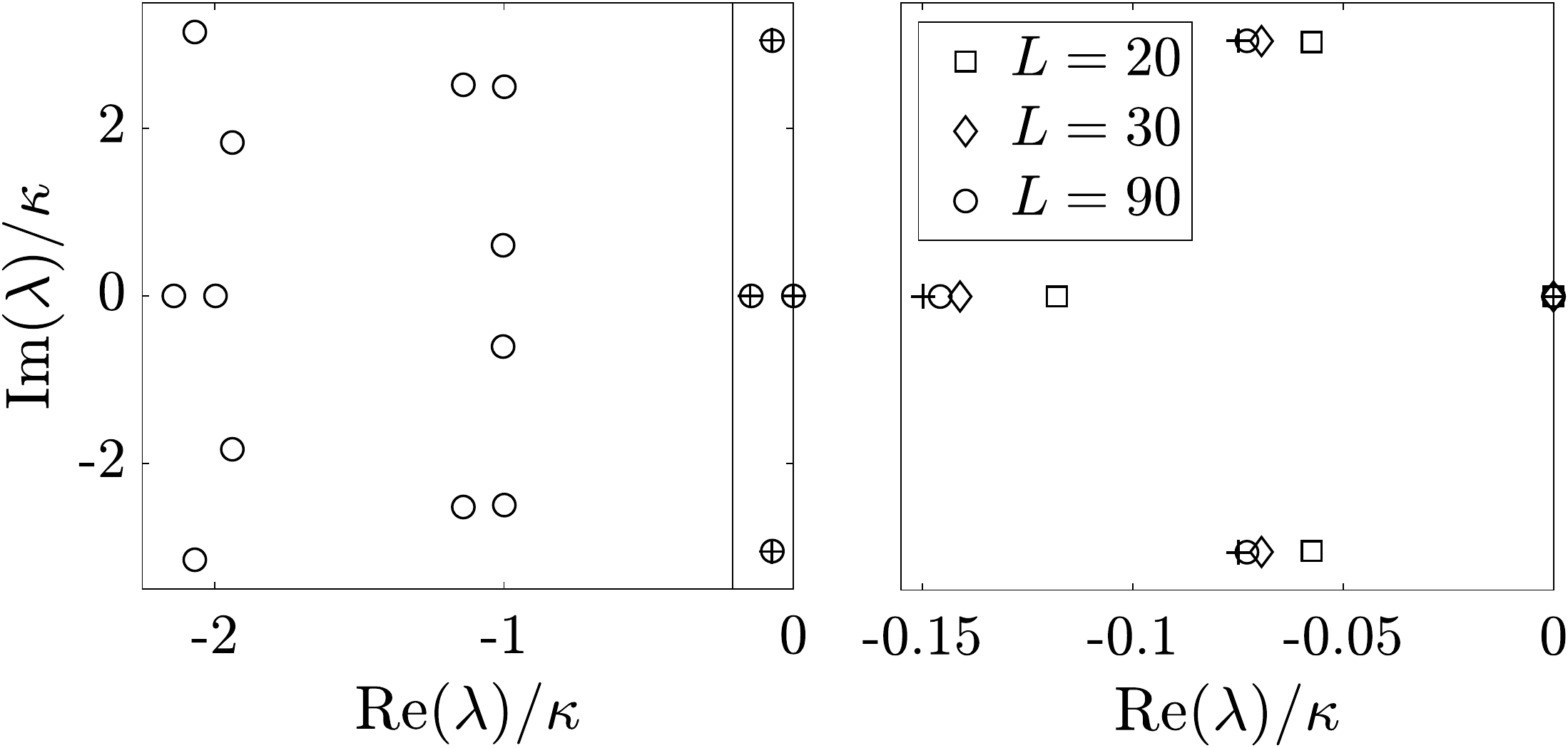}\vspace{-3ex}
		\flushleft\normalsize{(b)}\hspace{-2.4ex}\includegraphics[width=\linewidth]{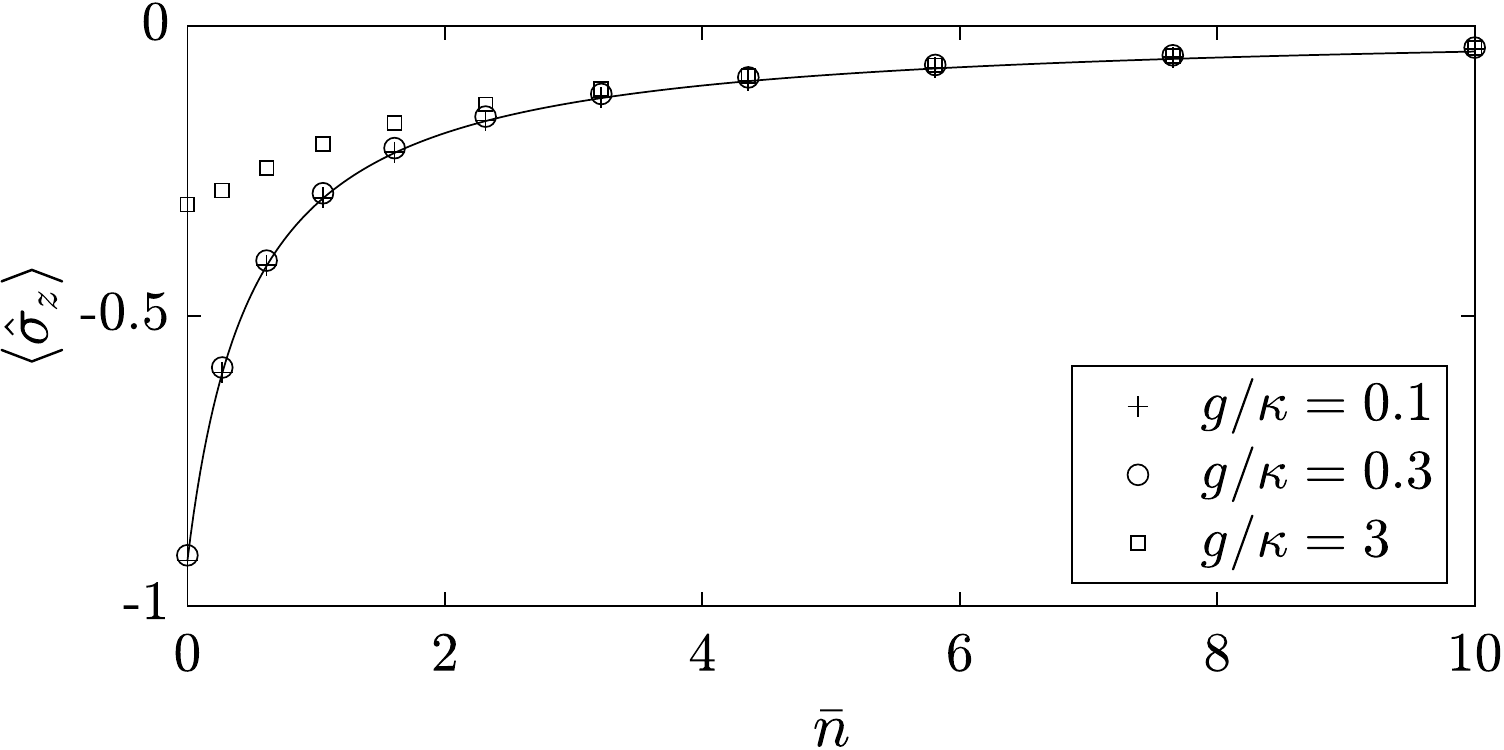}
		\caption{(a) Left: Eigenvalue spectra of $\mathcal{L}$ (circles) and $\mathcal{L}_\text{eff}$ (pluses). Right: Zoom of the four eigenvalues of $\mathcal{L}$ that correspond to the effective spectrum for different truncations $L$ of the Hilbert-space dimension. (b) Steady-state expectation value $\langle \hat{\sigma}^z\rangle$ as a function of $\bar{n}$ for $\omega_c/\kappa=2.5$ and $\omega_0/\kappa=3$. The solid line is the effective result and markers represent the results of the full master equation for $g/\kappa=0.1,0.3,3$.\label{Fig:4}}
	\end{figure}
	In Fig.~\ref{Fig:4}, we compare the eigenvalue spectra of the full Liouvillian $\mathcal{L}$ and the effective one $\mathcal{L}_\text{eff}$. The parameters are $\omega_c/\kappa=2.5$, $\omega_0/\kappa=3$, $g/\kappa=0.1$, and $\bar{n}=4$. In the left panel, we find good agreement of the four eigenvalues of $\mathcal{L}_\text{eff}$ with eigenvalues of the exact spectrum. The right panel represents a zoom of the area to the right of the vertical line in the left panel, which contains these four eigenvalues. The different markers correspond to different truncation dimensions $L$ of the BM Hilbert-space for the numerical simulation of the full master equation. We find that even for a relatively low thermal occupation number, $\bar{n}=4$ in this case, which roughly corresponds, for example, to a BM frequency of 10~GHz at a temperature of 2~K, Hilbert-space dimensions of $L=20$ are not sufficient to provide accurate numerical results and we have to resort to $L>80$ to attain a satisfactory convergence. This implies that we reduce the underlying Liouville space dimension by a factor of $80^2=6400$ which highlights the efficiency of our effective approach. 
	
	In the effective description, the atomic steady state can be calculated straightforwardly from $\mathcal{L}_\text{eff}$ and has the form
	\begin{equation}
		\label{eq:rabi_sz}
		\hat{\rho}_\text{st}=\frac{1}{2}-\frac{\omega_c\omega_0}{(\omega_c^2+\omega_0^2+\kappa^2)(2\bar{n}+1)}\hat{\sigma}^z.
	\end{equation}
	Figure~\ref{Fig:4}(b) shows the steady-state expectation value $\langle \hat{\sigma}^z\rangle$ as a function of the thermal occupation number $\bar{n}$ of the BM. The solid line represents the effective result $\langle \hat{\sigma}^z\rangle=-2\omega_c\omega_0/[(\omega_c^2+\omega_0^2+\kappa^2)(2\bar{n}+1)]$, obtained from Eq.~\eqref{eq:rabi_sz}, and the markers represent the results of the full master equation for different values of the coupling constant $g$. The remaining parameters are the same as in Fig.~\ref{Fig:4}(a), and the three values of the coupling strength are $g/\kappa=0.1,0.3,3$. For these values of $g$ we respectively find $||\hat{\alpha}||=g/\sqrt{(\omega_c-\omega_0)^2+\kappa^2}\approx 0.09,0.27,2.68$ for the perturbation parameter of the QS-BM decoupling. For $g/\kappa=0.1$ (pluses) the exact results agree well with the effective one and even for $g/\kappa=0.3$ (circles) they overlap to a good degree. Only for a strong coupling $g>\kappa$ [e.g., the squares in Fig.~\ref{Fig:2}(b) for $g/\kappa=3$] do we start to see clear discrepancies. Convergence tests show that, in this case, for the calculation of $\langle \hat{\sigma}^z \rangle$, only $L\gtrsim65$ is sufficient for numerical simulations using $\mathcal{L}$.

	\section{Conclusions}
	\label{Sec:conclusion}
	In this paper, we have applied an effective Lindblad master equation to describe the cooling dynamics of a quantum system that is coupled to a dissipative bosonic mode. We showed that this approach can correctly describe the unresolved and resolved sideband-cooling regime in an optomechanical setup. We compared our approach to a numerical treatment of mechanical oscillator plus bosonic mode as well as to previous results obtained for weak light-matter coupling. Remarkably, we found that the effective approach can also describe the correct steady state for rather large values of the light-matter coupling. As a next step, we studied the cooling dynamics of an interacting many-body quantum system described by the transverse-field Ising model, which is coupled to a bosonic mode. We were able to show that by understanding the spectral properties of the transverse-field Ising model we can cool the system into its ground state. We described this cooling method effectively by simulating only the dynamics of the spins  and compared it to a full simulation of the spins and bosonic mode. We found excellent agreement highlighting the possibility to use the effective master equation to also describe cooling of many-body quantum systems. Finally, we generalized this effective master equation to be applicable also in the regime where the bosonic mode has a nonvanishing thermal occupation. With this new master equation we have described a single spin coupled to a thermal and dissipative bosonic mode. Here, we compared the steady state of the spin obtained from the effective description to the steady state of the composite dynamics of the spin and field. We found very good agreement even for intermediate coupling strengths. Most significantly, we were able to describe the correct frequency shifts and damping rates of the spin without including the bosonic mode, which could otherwise be adequately modeled only with a very large Fock-space cutoff. This highlights the tremendous reduction of the Liouville-space dimension that we achieve by using the effective Lindblad master equation. 
	
	In this paper we focused on a comparison of the effective Lindblad master equation techniques to existing and exact results. As a next step, one could model with this Lindblad master equation various other cooling techniques that are more evolved or for which efficient numerical approaches are missing. In the future, we aim to also include dissipation of the quantum system itself in the description, which could significantly modify its dynamics and possibly the cooling efficiency. Additionally, it would be interesting to see the effect that this dissipation has on the elimination of the bosonic modes. Furthermore, we believe that there is great potential of this theory to also describe collective decay mechanisms, which could be used to engineer superradiant and subradiant states that might be useful both for cooling protocols~\cite{Chan:2002,Domokos:2002,Black:2003,Xu:2016,Jaeger:2017} and for the preparation of entangled many-body states~\cite{DallaTorre:2013,Reilly:2022}.

	\section*{Acknowledgments}
	We thank C. A. Gonz\'alez-Guti\'errez and J. H. Zhang for helpful discussions.
	S.B.J. is supported by the Research Centers of the Deutsche Forschungsgemeinschaft (DFG), Projects No.~A4 and No.~A5 in SFB/Transregio 185: “OSCAR". R.B. acknowledges start-up funding of the Huazhong University of Science and Technology.

	\appendix
	
	\section{Derivation of the effective master equation with a thermal environment}
	\label{app:A}
	The procedure for the elimination of the BM at zero temperature and the ensuing effective master equation were reported in Ref.~\cite{Jaeger:2022}. In this section, we present the details of the modifications of those results that are due to a environment with non-vanishing thermal occupation $\bar{n}$.
	
	Like for the zero-occupation limit, we use a displacement operator that is essentially a generalization of the polaron transformation~\cite{Mahan2000} and has the form
	\begin{equation}
		\hat{D}=\exp[\hat{a}^{\dag}\hat{\alpha}(t)-\hat{\alpha}^{\dag}(t)\hat{a}]\label{D},
	\end{equation}
	where we have introduced the effective-field operator $\hat{\alpha}(t)$. We apply this transformation onto the master equation and define the displaced density operator as
	\begin{equation}
		\label{eq:disrho}
		\tilde{\rho}=\hat{D}^{\dag}\hat{\rho}\hat{D}.
	\end{equation}
	Now, the idea is that one optimizes $\hat{\alpha}$ in order to decouple the BM from the QS and such that the density operator of the BM is to good approximation in a thermal state [Eq.~\eqref{eq:thermal}]. To achieve this, we assume $\hat{S}$ and the effective-field $\hat{\alpha}$ are sufficiently small and perform a second-order perturbation theory for these operators. 
	
	We follow the steps of Ref.~\cite{Jaeger:2022} and write down the dynamics of $\tilde{\rho}$ in the form
	\begin{equation}
		\label{eq:generalmaster}
		\frac{\partial\tilde{\rho}}{\partial t}=\mathcal{L}_a\tilde{\rho}+\mathcal{L}_b\tilde{\rho}.
	\end{equation}
	The first term is given by
	\begin{equation}
		\mathcal{L}_a\tilde{\rho}=\frac{\partial \hat{D}^{\dag}}{\partial t}\hat{D}\tilde{\rho}+\tilde{\rho}\hat{D}^{\dag}\frac{\partial \hat{D}}{\partial t},   
	\end{equation} 
	and originates from a possible explicit time dependence of $\hat{\alpha}$. The second term , on the other hand, originates from the displaced Lindbladian and reads
	\begin{align}
		\mathcal{L}_b\tilde{\rho}=\hat{D}^{\dag}\frac{\partial \hat{\rho}}{\partial t}\hat{D}=&-i[\hat{D}^{\dag}\hat{H}\hat{D},\tilde{\rho}]\nonumber\\
		&+\kappa{(\bar{n}+1)\tilde{\mathcal{D}}[\hat{a}]\tilde{\rho}+\kappa\bar{n}\tilde{\mathcal{D}}[\hat{a}^{\dag}]}\tilde{\rho}\label{Displacedmaster}.
	\end{align}
	Here, we have defined the displaced Hamiltonian $\hat{D}^{\dag}\hat{H}\hat{D}$ as well as the  dissipators $\tilde{\mathcal D}[\hat{a}]\tilde{\rho}=\hat{D}^{\dag}(\mathcal{D}[\hat{a}]\hat{\rho})\hat{D}$, and $\tilde{\mathcal D}[\hat{a}^{\dag}]\tilde{\rho}=\hat{D}^{\dag}(\mathcal{D}[\hat{a}^{\dag}]\hat{\rho})\hat{D}$. In the following, we will give the explicit forms of the two superoperators $\mathcal{L}_a$ and $\mathcal{L}_b$.
	
	\subsection{Calculation of $\mathcal{L}_a$}
	\label{App:A:1}
	Since a finite occupation of the BM does not modify the $\mathcal{L}_a$ term, we simply report the final result obtained in Ref.~\cite{Jaeger:2022}. This result reads
	\begin{equation}
		\mathcal{L}_a\tilde{\rho}=-i\bigg[-i\hat{D}^{\dag}\frac{\partial \hat{D}}{\partial t},\tilde{\rho}\bigg]
	\end{equation}
	with
	
		\begin{widetext}
	\begin{align}
		\hat{D}^{\dag}\frac{\partial \hat{D}}{\partial t}=\bigg(\hat{a}^\dag\frac{\partial\hat{\alpha}}{\partial t}-\frac{\partial\hat{\alpha}^{\dag}}{\partial t}\hat{a}\bigg)
		+\frac{1}{2}\bigg(\hat{\alpha}^\dag\frac{\partial\hat{\alpha}}{\partial t}-\frac{\partial\hat{\alpha}^\dag}{\partial t}\hat{\alpha}
		-\bigg[\frac{\partial \hat{\alpha}^{\dag}}{\partial t},\hat{r}\bigg]\hat{a}-\hat{a}^{\dag}\bigg[\hat{r},\frac{\partial\hat{\alpha}}{\partial t}\bigg]\bigg), \label{DdotD2}
	\end{align}
	where we have defined $\hat{r}=\hat{a}^\dag\hat{\alpha}-\hat{\alpha}^{\dag}\hat{a}$.

		\subsection{Calculation of $\mathcal{L}_b$}\label{App:A:2}
		When calculating $\mathcal{L}_b$ we proceed according to Ref.~\cite{Jaeger:2022} to find the displaced Hamiltonian
		\begin{equation}
			\hat{D}^{\dag}\hat{H}\hat{D}\approx\tilde{A}+\tilde{B}+\tilde{C}.
		\end{equation}
		The terms $\tilde{A}$, $\tilde{B}$, and $\tilde{C}$ are given up to second order in the operators $\hat{\alpha}$ and $\hat{S}$. The first term is the displaced Hamiltonian $\hat{H}_S$ and has the form
		\begin{align}
			\label{eq:A}\tilde{A}=\hat{D}^{\dag}\hat{H}_S\hat{D}\approx&\hat{H}_S+[\hat{\alpha}^{\dag},\hat{H}_S]\hat{a}+\hat{a}^{\dag}[\hat{H}_S,\hat{\alpha}]+\frac{1}{2}([\hat{\alpha}^{\dag},\hat{H}_S]\hat{\alpha}+\hat{\alpha}^{\dag}[\hat{H}_S,\hat{\alpha}])\nonumber\\
			&+\frac{1}{2}(\bm{[}\hat{\alpha}^{\dag}\hat{a}-\hat{a}^{\dag}\hat{\alpha},[\hat{\alpha}^{\dag},\hat{H}_S]\bm{]}\hat{a}+\hat{a}^{\dag}\bm{[}\hat{\alpha}^{\dag}\hat{a}-\hat{a}^{\dag}\hat{\alpha},[\hat{H}_S,\hat{\alpha}]\bm{]}).
		\end{align}
		The $\hat{B}$ term is given by
		\begin{align}\label{eq:B}
			\tilde{B}=\hat{D}^{\dag}\hat{a}^{\dag}\hat{\Omega}_S\hat{a}\hat{D}
			\approx&\hat{a}^{\dag}\hat{\Omega}_S\hat{a}+\hat{\alpha}^{\dag}\hat{\Omega}_S\hat{a}+\hat{a}^{\dag}\hat{\Omega}_S\hat{\alpha}+\hat{\alpha}^{\dag}\hat{\Omega}_S\hat{\alpha}\nonumber\\
			&+\hat{a}^{\dag}([\hat{\alpha}^{\dag},\hat{\Omega}_S]\hat{a}+\hat{a}^{\dag}[\hat{\Omega}_S,\hat{\alpha}])\hat{a}\nonumber+\frac{1}{2}\hat{a}^{\dag}([\hat{\alpha}^{\dag},\hat{\Omega}_S]\hat{\alpha}+\hat{\alpha}^{\dag}[\hat{\Omega}_S,\hat{\alpha}])\hat{a}\nonumber\\
			&+\hat{\alpha}^{\dag}([\hat{\alpha}^{\dag},\hat{\Omega}_S]\hat{a}+\hat{a}^{\dag}[\hat{\Omega}_S,\hat{\alpha}])\hat{a}+\frac{1}{2}(\hat{a}^{\dag}[\hat{\alpha}^{\dag},\hat{\alpha}]-[\hat{\alpha}^{\dag},\hat{\alpha}^{\dag}]\hat{a})\hat{\Omega}_S\hat{a}\nonumber\\
			&+\hat{a}^{\dag}([\hat{\alpha}^{\dag},\hat{\Omega}_S]\hat{a}+\hat{a}^{\dag}[\hat{\Omega}_S,\hat{\alpha}])\hat{\alpha}+\frac{1}{2}\hat{a}^{\dag}\hat{\Omega}_S([\hat{\alpha}^{\dag},\hat{\alpha}]\hat{a}-\hat{a}^{\dag}[\hat{\alpha},\hat{\alpha}])\nonumber\\
			&+\frac{1}{2}\hat{a}^{\dag}(\bm{[}\hat{\alpha}^{\dag}\hat{a}-\hat{a}^{\dag}\hat{\alpha},[\hat{\alpha}^{\dag},\hat{\Omega}_S]\bm{]}\hat{a}+\hat{a}^{\dag}\bm{[}\hat{\alpha}^{\dag}\hat{a}-\hat{a}^{\dag}\hat{\alpha},[\hat{\Omega}_S,\hat{\alpha}]\bm{]})\hat{a}
		\end{align}
		and the third term, $\hat{C}$, is found to read
		\begin{align}
			\label{eq:C}
			\tilde{C}=\hat{D}^{\dag}[\hat{a}^{\dag}\hat{S}+\hat{S}^{\dag}\hat{a}]\hat{D}\approx&\hat{a}^{\dag}\hat{S}+\hat{S}^{\dag}\hat{a}+\hat{\alpha}^{\dag}\hat{S}+\hat{S}^{\dag}\hat{\alpha}\nonumber\\
			&+\hat{a}^{\dag}([\hat{\alpha}^{\dag},\hat{S}]\hat{a}+\hat{a}^{\dag}[\hat{S},\hat{\alpha}])+([\hat{\alpha}^{\dag},\hat{S}^{\dag}]\hat{a}+\hat{a}^{\dag}[\hat{S}^{\dag},\hat{\alpha}])\hat{a}.
		\end{align}
		At this point, the main difference from the derivation in Ref.~\cite{Jaeger:2022} is that there is an additional dissipator. For the first term, we obtain the same result, which is
		\begin{align}
			\tilde{\mathcal{D}}[\hat{a}]\tilde{\rho}=&	\hat{D}^{\dag}(2\hat{a}\hat{\rho}\hat{a}^{\dag}-\hat{a}^{\dag}\hat{a}\hat{\rho}-\hat{\rho}\hat{a}^{\dag}\hat{a})\hat{D}\nonumber\\
			\approx&2\hat{a}\tilde{\rho}\hat{a}^{\dag}-\hat{a}^{\dag}\hat{a}\tilde{\rho}-\tilde{\rho}\hat{a}^{\dag}\hat{a}+2\hat{a}\tilde{\rho}\hat{\alpha}^{\dag}-\hat{\alpha}^{\dag}\hat{a}\tilde{\rho}-\tilde{\rho}\hat{\alpha}^{\dag}\hat{a}+2\hat{\alpha}\tilde{\rho}\hat{a}^{\dag}-\hat{a}^{\dag}\hat{\alpha}\tilde{\rho}-\tilde{\rho}\hat{a}^{\dag}\hat{\alpha}\nonumber\\
			&+2\hat{\alpha}\tilde{\rho}\hat{\alpha}^{\dag}-\hat{\alpha}^{\dag}\hat{\alpha}\tilde{\rho}-\tilde{\rho}\hat{\alpha}^{\dag}\hat{\alpha}+\{([\hat{\alpha}^{\dag},\hat{\alpha}]\hat{a}-\hat{a}^{\dag}[\hat{\alpha},\hat{\alpha}])\tilde{\rho}\hat{a}^{\dag}+\text{H.c.}\}\nonumber\\
			&-\frac{1}{2} \{\hat{a}^{\dag}([\hat{\alpha}^{\dag},\hat{\alpha}]\hat{a}-\hat{a}^{\dag}[\hat{\alpha},\hat{\alpha}])\tilde{\rho}+\text{H.c.}\}-\frac{1}{2}\{\tilde{\rho}\hat{a}^{\dag}([\hat{\alpha}^{\dag},\hat{\alpha}]\hat{a}-\hat{a}^{\dag}[\hat{\alpha},\hat{\alpha}])+\text{H.c.}\}.\label{Diss1}
		\end{align}
		The new term that originates from thermal excitations is given by
		\begin{align}
			\tilde{\mathcal{D}}[\hat{a}^{\dag}]\tilde{\rho}=&	\hat{D}^{\dag}(2\hat{a}^\dag\hat{\rho}\hat{a}-\hat{a}\hat{a}^\dag\hat{\rho}-\hat{\rho}\hat{a}\hat{a}^\dag)\hat{D}\nonumber\\
			\approx&2\hat{a}^\dag\tilde{\rho}\hat{a}-\hat{a}\hat{a}^\dag\tilde{\rho}-\tilde{\rho}\hat{a}\hat{a}^\dag+2\hat{\alpha}^\dag\tilde{\rho}\hat{a}-\hat{a}\hat{\alpha}^\dag\tilde{\rho}-\tilde{\rho}\hat{a}\hat{\alpha}^\dag+2\hat{a}^\dag\tilde{\rho}\hat{\alpha}-\hat{\alpha}\hat{a}^\dag\tilde{\rho}-\tilde{\rho}\hat{\alpha}\hat{a}^\dag\nonumber\\
			&+2\hat{\alpha}^\dag\tilde{\rho}\hat{\alpha}-\hat{\alpha}\hat{\alpha}^\dag\tilde{\rho}-\tilde{\rho}\hat{\alpha}\hat{\alpha}^\dag+\{\hat{a}^{\dag}\tilde{\rho}([\hat{\alpha}^{\dag},\hat{\alpha}]\hat{a}-\hat{a}^{\dag}[\hat{\alpha},\hat{\alpha}])+\text{H.c.}\}\nonumber\\
			&-\frac{1}{2} \{([\hat{\alpha}^{\dag},\hat{\alpha}]\hat{a}-\hat{a}^{\dag}[\hat{\alpha},\hat{\alpha}])\hat{a}^{\dag}\tilde{\rho}+\text{H.c.}\}-\frac{1}{2}\{\tilde{\rho}([\hat{\alpha}^{\dag},\hat{\alpha}]\hat{a}-\hat{a}^{\dag}[\hat{\alpha},\hat{\alpha}])\hat{a}^{\dag}+\text{H.c.}\}.\label{Diss2}
		\end{align}\newpage
			\end{widetext}

	\section{Projecting on the thermal state}
	\label{app:B}
	We will now assume that 
	\begin{equation}
		\tilde{\rho}=\hat{\rho}_{\mathrm{sys}}\otimes\hat{\rho}_{\mathrm{th}}+\hat{\xi}
	\end{equation}
	such that $\hat{\xi}$ is a traceless operator 
	and at least of second order in perturbation theory,
	 $\hat{\rho}_{\mathrm{sys}}$ is the density operator describing the QS, and $\hat{\rho}_{\mathrm{th}}$ is the thermal state of the BM given by Eq.~\eqref{eq:thermal}.

	In Ref.~\cite{Jaeger:2022}, the authors were able to find a good choice for the $\hat\alpha$ by simply collecting whenever $\hat{a}^{\dag}$ was operating on $\tilde{\rho}$ from the left. This was possible since it was assumed that $\hat{\rho}_{\mathrm{th}}$ had quasi-zero occupation, and therefore, they could neglect the left-operation of $\hat{a}$ on $\tilde{\rho}$. However, this is not possible for nonvanishing occupation numbers. 
	
	In this more general case, we first collect all operators in Eq.~\eqref{eq:generalmaster} that are of first order in the operators $\hat{S}$ and $\hat{\alpha}$. Setting the combination of all these terms to zero will result in a condition for $\alpha$ which minimizes the coupling between the BM and the QS. We will follow this procedure first by collecting all first order terms in Eqs.~\eqref{eq:A},~\eqref{eq:B}, and~\eqref{eq:C} in
	\begin{align}\label{eq:K0}
		\hat{K}_0=[\hat{E}_0^\dag\hat{a}+\hat{a}^\dag\hat{E}_0+\hat{F}_0^\dag\hat{a}^\dag\hat{a}\hat{a}+\hat{a}^\dag\hat{a}^\dag\hat{a}\hat{F}_0,\tilde{\rho}].
	\end{align}
	In this equation, we have collected linear terms in $\hat{a}$ and $\hat{a}^\dag$ in
	\begin{equation}
		\label{eq:E0}
		\hat{E}_0=-\frac{\partial\hat{\alpha}}{\partial t}-i[\hat{H}_S,\hat{\alpha}]-i\hat{\Omega}_S\hat{\alpha}-i\hat{S}.
	\end{equation}
	In addition, we have also found cubic contributions that are multiplied by
	\begin{align}
		\label{eq:F0}
		\hat{F}_0=-i[\hat\Omega_S,\hat{\alpha}].
	\end{align}
	These cubic terms do not allow us to find a single QS operator $\hat{\alpha}$ in order to achieve a vanishing first-order contribution. Therefore, in order to be able to drop the term~\eqref{eq:F0} we have to make an additional assumption to ensure that $\hat{F}_0$ is sufficiently small. This is the case if
	\begin{align}\label{eq:smallF}
		\|[\hat{\Omega}_S,\hat{\alpha}]\|\ll \|\Omega_S+i\kappa\|\|\alpha\|.
	\end{align}
	With this property fulfilled, we can assume that $\hat{F}_0$ is of higher order and can be neglected when calculating $\hat{\alpha}$. 
	
	Now, in the next step we want to collect all first-order terms that originate from the dissipator. To do this, we use the convention that whenever the Hamiltonian is already multiplied by the QS operator $\hat{\alpha}$ from the left, we shift all bosonic operators to the left of the density operator. This is enabled by explicitly using the fact that the BM are in a thermal state [Eq.~\eqref{eq:thermal}] and using the relation
	\begin{equation}
		\hat{\rho}_{\mathrm{th}}\hat{a}^{\dag}=\frac{\bar{n}}{\bar{n}+1}\hat{a}^{\dag}\hat{\rho}_{\mathrm{th}}.
	\end{equation}
	Collecting all QS operators that are in front of $\hat{a}^\dag$ and multiplied from the left to the QS operator we obtain for the expression in Eq.~\eqref{Diss1}
	\begin{equation}
		\hat{E}_{1}=\kappa(\bar{n}-1)\hat{\alpha}.
	\end{equation}
	Using the same method for Eq.~\eqref{Diss2} we find
	\begin{equation}
		\hat{E}_{2}=-\kappa\bar{n}\hat{\alpha}.
	\end{equation}
	Adding those two terms yields $\hat{E}_{1}+\hat{E}_{2}=-\kappa\hat{\alpha}$, and collecting all first-order terms we can rewrite them as a single commutator,
	\begin{equation}
		\hat{K}=[\hat{E}^\dag\hat{a}+\hat{a}^{\dag}\hat{E},\tilde{\rho}],    
	\end{equation}
	with $\hat{E}=\hat{E}_{0}+\hat{E}_{1}+\hat{E}_{2}$, which can be written as
	\begin{equation}
		\hat{E}=-\frac{\partial\hat{\alpha}}{\partial t}-i[\hat{H}_S,\hat{\alpha}]-i\hat{\Omega}_S\hat{\alpha}-i\hat{S}-\kappa\hat{\alpha}.
	\end{equation}
	
	With the help of this equation we now choose $\hat{\alpha}$ such that $\hat{E}=0$. Inserting this $\hat{\alpha}$ in the master equation and tracing over the BM degrees of freedom we find the master equation~\eqref{eq:Leff}. For completeness, we also report the effective Hamiltonian which can be directly calculated from Eqs.~\eqref{eq:A},~\eqref{eq:B}, and~\eqref{eq:C} and has the form
	\begin{align}
		\hat{H}_{\mathrm{eff}}=&\hat{H}_S-\frac{i}{2}\left(\hat{\alpha}^\dag\frac{\partial\hat{\alpha}}{\partial t}-\frac{\partial\hat{\alpha}^\dag}{\partial t}\hat{\alpha}\right)+\hat{\alpha}^{\dag}\hat{\Omega}_S\hat{\alpha}\nonumber\\
		&+\frac{1}{2}([\hat{\alpha}^{\dag},\hat{H}_S]\hat{\alpha}+\hat{\alpha}^{\dag}[\hat{H}_S,\hat{\alpha}])+\hat{\alpha}^{\dag}\hat{S}+\hat{S}^{\dag}\hat{\alpha}\nonumber\\
		&+\frac{i\bar{n}}{2}\left(\left[\frac{\partial \hat{\alpha}^{\dag}}{\partial t},\hat{\alpha}\right]-\left[\hat{\alpha}^\dag,\frac{\partial \hat{\alpha}}{\partial t}\right]\right)\nonumber\\
		&+\frac{\bar{n}}{2}(\bm{[}[\hat{\alpha}^\dag,\hat{H}_S],\hat{\alpha}\bm{]}+\bm{[}\hat{\alpha}^\dag,[\hat{H}_S,\hat{\alpha}]\bm{]})+\bar{n}\hat{\Omega}_S.\nonumber\\
	\end{align}
	In this expression, we have explicitly used assumption~\eqref{eq:smallF}. Now, using $\hat{E}=0$, we finally arrive at Eq.~\eqref{eq:Heff} of the main text.


%

\end{document}